# Quantum Computing Spacetime

P. A. Zizzi

Dipartimento di Astronomia  dell' Università di Padova
Vicolo dell' Osservatorio, 2
35122 Padova, Italy
zizzi@.pd.astro.it

## Abstract

A causal set C can describe a discrete spacetime, but this discrete spacetime is not quantum, because C is endowed with Boolean logic, as it does not allow cycles.
In a quasi-ordered set Q, cycles are allowed. In this paper, we consider a subset QC of a quasi-ordered set Q, whose elements are all the cycles.
In QC, which is endowed with quantum logic, each cycle of maximal outdegree N in a node, is associated with N entangled qubits. Then QC describes a quantum computing spacetime. This structure, which is non-local and non-casual, can be understood as a proto-spacetime.
Micro-causality and locality can be restored in the subset U of Q whose elements are unentangled qubits which we interpret as the states of quantum spacetime. The mapping of quantum spacetime into proto-spacetime is given by the action of the XOR gate.
Moreover, a mapping is possible from the Boolean causal set into U by the action of the Hadamard gate.
In particular, the causal order defined on the elements of U induces the causal evolution of spin networks.

# 1   Introduction

Discreteness of spacetime at the Planck scale seems to be one of the most compelling requirements of quantum gravity [1], the theory which should reconcile and unify Quantum Mechanics and General Relativity. In fact, both loop quantum gravity [2] and superstrings/M-theory [3], the two major candidates for quantum gravity, strongly suggest that spacetime at the Planck scale must have a discrete structure.
In particular, in loop quantum gravity, non-perturbative techniques have led to a picture of quantum geometry, which is rather of a polymer type, and geometrical quantities such area and volume have discrete spectra. In quantum geometry, spin networks play a very important role. They were invented by Penrose [4] and lead to a drastic change in the concept of space-time, going from that of a smooth manifold to that of a discrete, purely combinatorial structure. Then, spin networks were re-discovered by Rovelli and Smolin [5] in the context of loop quantum gravity, where they are eigenstates of the area and volume-operators [6].
However, this theory of quantum geometry, does not reproduce classical General Relativity in the continuum limit. Recent models of quantum gravity called "spin foam models" [7] seem to have continuum limits.
Anyway, as spin foam models are Euclidean, they are not suitable to recover causality at the Planck scale. For this purpose, the theory should be intrinsically Lorentzian. However, the very concept of causality becomes uncertain at the Planck scale, when the metric undergoes quantum fluctuations as Penrose [8] argued. So, one should consider a discrete alternative to the Lorentzian metric, which is the causal set (a partially ordered set-or poset- whose elements are events of a discrete space-time). Such theories of quantum gravity based on the casual set were formulated by Sorkin et al. [9].
Rather recently, a further effort in trying to recover causality at the Planck scale, has been undertaken by Markopoulou and Smolin [10]. They considered the evolution of spin networks in discrete time steps, and they claimed that the evolution is causal because the history of evolving spin networks is a causal set.
 Of course, also the causal set approach [9] to quantum gravity, relies on a discrete structure of spacetime at the fundamental level.
Finally, the quantum computational approach to quantum gravity [11] suggests as well that spacetime is discrete at the Planck scale. This fourth approach has been applied in particular to quantum cosmology, resulting in a model of quantum inflation describing the very early universe as a growing quantum network [12].
In this paper, we will investigate the structure of quantum spacetime by applying the tools of quantum information and quantum computation [13] to an extended version of the causal set theory. Actually, we will not consider the partially ordered set (poset) on which the causal set C is based, but the quasi ordered set Q, which allows closed loops. The poset was choosen to describe a discrete spacetime with micro-causality, just because a poset does not allows cycles. However, we show that this restriction is in disagreement with the very nature of quantum spacetime which should be endowed with quantum logic. On the contrary, a poset, is endowed with Boolean logic. We interpret the one cycle graph in Q as the one-qubit state (or quantum bit, the unit of quantum information). There are two important subsets of Q, one whose elements are entangled qubits (which we call QC, i.e., quantum computing spacetime), and one whose elements are unentangled qubits (which we call U). In QC micro-causality is missing, as well as locality. Instead, we show that in U, it is possible to define a causal order. This is due to the fact that unentangled qubits are product states, then it is possible to define an increase of information entropy which induces an arrow of discrete time. We interpret QC as a proto-spacetime, while U plays the role of quantum spacetime itself. The elements of U are qubits, i.e., superposed states, then U is endowed with quantum logic. The elements of U are the "quantum events". Instead, in the Boolean causal set C considered by Sorkin and coworkers [9] the events of discrete spacetime are points and not cycles. However, we show that the causal set C (or, better, its subset B-where B stands for Boolean- whose elements are classical bits) can be mapped into the set U by a quantum logic gate (the Hadamard gate).

Also, the mapping from U to the proto-spacetime QC is made by the XOR gate.

In summary, spacetime at the fundamental level, shows a quite rich spectrum of different structures which can be mapped into each other by quantum logic gates.

Finally, we show that there is a one-to-one relation between the "quantum events" (elements of U) and the punctures of spin networks' edges. In fact, the elements of U can be interpreted as pairs of virtual events which are the birth and death of a Planckian black hole, which has a horizon area of one pixel (one unit of Planck area). By the quantum version [11] of the holographic principle [14], each pixel of area encodes one qubit. Moreover, we know from loop quantum gravity, that if a 2-surface is punctured by a spin network edge in one point, it acquires an area of one pixel. So, each (extended) quantum event corresponds to one point of discrete spacetime, i.e, to one puncture of a spin networks' edge. Then, the causal relation defined on the elements of U induces the causal evolution of spin networks.

## 2. A brief review of ordered sets

### 2.1 Partially ordered set P

A partially ordered set (or poset) **P** is a set S plus a relation $\leq$ on the set, with the following properties:
1. Reflexivity: $a \leq a$ for all $a \in S$
2. Antisymmetry: $a \leq b$ and $b \leq a$ implies $a = b$
3. Transitivity: $a \leq b$ and $b \leq c$ implies $a \leq c$

### 2.2 Totally ordered set T

A totally ordered set **T** is a set S plus a relation $R$ on the set called total order, that satisfies the conditions for a partial order plus the comparability condition (or trichotomy law):
1. Reflexivity: $a \leq a$ for all $a \in S$
2. Antisymmetry: $a \leq b$ and $b \leq a$ implies $a = b$
3. Transitivity: $a \leq b$ and $b \leq c$ implies $a \leq c$
4. Comparability: either $a \leq b$ or $b \leq a$ for any $a,b \in S$

### 2.3 Quasi-ordered set Q

A quasi-ordered set **Q** is a set S plus a relation $\leq$ on the set, which satisfies the properties of reflexivity and transitivity:
1. Reflexivity: $a \leq a$ for all $a \in S$
3. Transitivity: $a \leq b$ and $b \leq c$ implies $a \leq c$

A quasi-order does not satsfy antisymmetry, so cycles are allowed in **Q.**
In **Q** we can have pairs (a,b) of three different types:
i) incomparable: neither $a \leq b$ nor $b \leq a$
ii) comparable: either $a \leq b$ or $b \leq a$ ($a \leq b$ and $b \leq a$ implies $a = b$)
iii) comparable but not equivalent: $a \leq b$ and $b \leq a$ ($a \leq b$ and $b \leq a$ with $a \neq b$)
Of course, any poset is also quasi-ordered.

## 3. The causal set C

A causal set C is a locally finite, partially ordered set, whose elements are events of a discrete space-time.
For a causal set C, the following properties hold:
1. Reflexivity: $p \leq p$ for all $p \in C$
2. Antisymmetry: $p \leq q$ and $q \leq p$ implies $p = q$
3. Transitivity: $p \leq q$ and $q \leq r$ implies $p \leq r$
4. Local finiteness: $|A(p,q)| < \infty$

Where $|A(p,q)|$ is the cardinality of the "Alexandrov set" $A(p,q)$ of two events p and q, which is the set of all events x such that $p \leq x \leq q$.
In particular, antisymmetry (or acyclicity) is needed to avoid closed timelike loops.
It is generally believed that the causal set C can describe a quantum spacetime endowed with micro-causality.
However, the underlying logic of the causal set C is classical, i.e, Boolean.
In fact, once we define: $p \leq q$ as "yes" and $q \leq p$ as "no", antisymmetry implies: either "yes" or "no". This means that the information stored in a causal set is given in terms of classical bits "0" and "1", as for example, in a classical computer.
However, if the aim is to describe a quantum spacetime, one should deal with a discrete structure whose underlying logic is quantum.

## 4. The quantum computing set QC

A digraph (or directed graph) is a graph in which each edge is replaced by a directed edge.
A digraph G is transitive if any three vertices a,b,c such that edges $(a,b),(b,c) \in G$ imply $(a,c) \in G$.
An oriented graph is a digraph having no symmetric pair of directed edges.
Moreover, a simple graph is a graph in which each pair of vertices are connected by at most one edge, while in a non-simple graph multiple edges are also permitted.
The indegree (outdegree) is the number of incoming (outgoing) directed edges in a node and the local degree is the total number of directed edges visiting a node.
A cycle: $p \leq q \leq p$, with $p \neq q$, in **Q,** implies "yes" and "no" at the same time, which is a non-Boolean proposition. The superposition of bits "0" and "1" is a quantum bit of information (or qubit).
The single qubit can be written as: $a|0\rangle + b|1\rangle$ where a and b are the complex amplitudes of the two states, with the condition: $|a|^2 + |b|^2 = 1$.
Then, some of the information stored in **Q** is then given in terms of qubits as in a quantum computer.

Given a quasi-ordered set **Q**, let us consider first only those pairs of elements (p,q) which are related but not equivalent (cycles): $p \leq q$ and $p \leq q$ with $p \neq q$. We will indicate this subset of **Q** as **QC**, where **QC** stands for "Quantum Computing".

Now, let us consider only those pairs of elements in **Q** which are related: either $p \leq q$ or $p \leq q$. We will indicate this subset of **Q** as **B**, where **B** stands for "Boolean". **B** is also a subset of a causal set **C**. All the "events" of B are the classical bits "0" and "1". The sets **B** and **QC** are disjoint: $B \cap QC = \emptyset$ where $B \subset C \subset Q$ and $QC \subset Q$.

In **Q** there are also pairs of elements which are cycles, but are not entangled with other cycles. For example, let us consder the cycle $p \leq q$ and $p \leq q$ with $p \neq q$, where one element of the cycle, let us say q, is related to a third element r:
$q \leq r$. This subset of **Q** will be called **U** where **U** stands for "unentangled".

In **U**, as it will be showed in the following, the concepts of time flow, micro-causality, and locality are still valid. The same concepts are instead completely lost in **QC**, which is to be considered just as a proto-structure of quantum spacetime.

The passage from the Boolean logic of **B** to the quantum logic of **QC** can be interpreted in terms of the action of two quantum logic gates, the Hadamard gate $H$, and the XOR gate (see fig. 1). In fact, first the Hadamard gate transforms the classical bits into superposed states (qubits), then the XOR gate transforms the superposed states into entangled states, as it will be showed in what follows.

The Hadamard gate is: $H = \frac{1}{\sqrt{2}}\begin{pmatrix} 1 & 1 \\ -1 & 1 \end{pmatrix}$. Its action on bits $|0\rangle$ and $|1\rangle$ gives a symmetric and an antisymmetric 1-qubit state respectively:

$H|0\rangle = \frac{1}{\sqrt{2}}(|1\rangle + |0\rangle)$ and $H|1\rangle = \frac{1}{\sqrt{2}}(|1\rangle - |0\rangle)$

Where we have represented the base states $|1\rangle$ and $|0\rangle$ as the vectors $\begin{pmatrix} 1 \\ 0 \end{pmatrix}$ and $\begin{pmatrix} 0 \\ 1 \end{pmatrix}$ respectively.

The quantum logic gate which transforms unentangled qubits (elements of **U**) into entangled qubits (elements of **QC**) is the XOR gate.

The XOR gate (or controlled-NOT gate) is the standard 2-qubits gate, and illustrates the interactions between two quantum systems.

Any quantum computation can be performed by using the XOR gate, and the set of one-qubit gates.

The XOR gate flips the "target" imput if its "control" imput is $|1\rangle$ and does nothing if it is $|0\rangle$:

$|1\rangle$--------------------        ----------------$|1\rangle$
                       XOR         i.e      $|10\rangle \rightarrow |11\rangle$

$|0\rangle$--------------------        ----------------$|1\rangle$

$|0\rangle$-----------------------        ---------------$|0\rangle$
                       XOR         i.e      $|00\rangle$
                                                        unchanged

$|0\rangle$-----------------------        ---------------$|0\rangle$

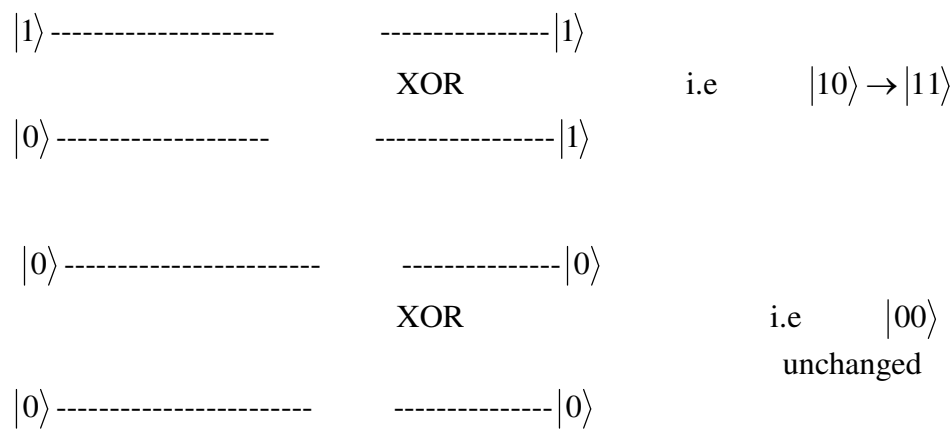

Hence, a XOR gate can clone Boolean imputs. But if one tries to clone a superposed state, one gets an entangled state:

$$\frac{1}{\sqrt{2}}(|0\rangle + |1\rangle) \text{------------}$$

$$\text{XOR} \text{-------------} \frac{1}{\sqrt{2}}(|00\rangle + |11\rangle)$$

(entangled state)

$$|0\rangle \text{--------------------------}$$

Then, the XOR gate cannot be used to copy superposed states (impossibility of cloning an unknown quantum state).

**Fig. 1**

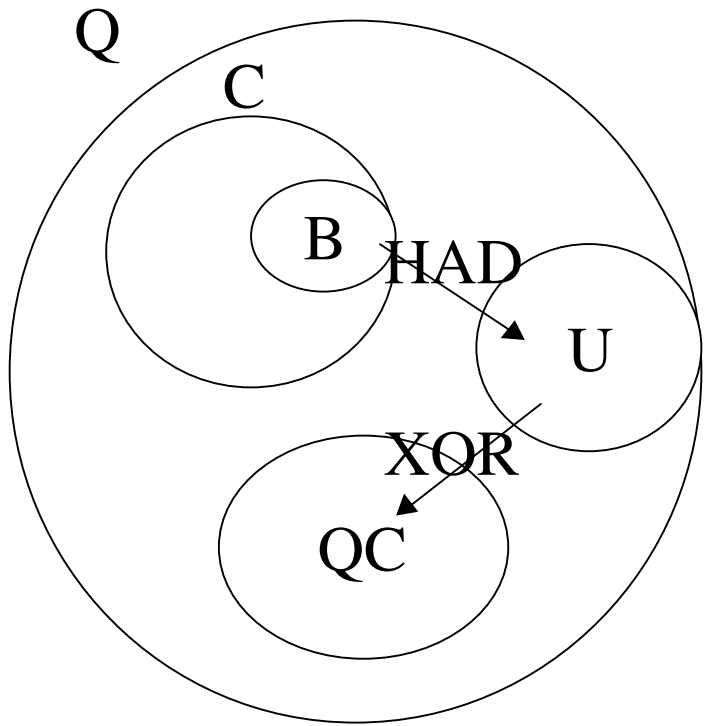

## 4  Cycles as qubits

Let us consider two "events" p and q in **QC**: $p \leq q \leq p$, with $p \neq q$.
This is a cycle graph, in paticular it is the $Z_2$ graph $\{0,1\}$ which is associated with the symmetric 1-qubit $|Q_1\rangle^S = \frac{1}{\sqrt{2}}(|0\rangle + |1\rangle)$ if the orientation is clockwise and with the antisymmetric 1-qubit $|Q_1\rangle^A = \frac{1}{\sqrt{2}}(|0\rangle - |1\rangle)$ if the orientation is anti-clockwise, as shown in fig.2.

**Fig.2**

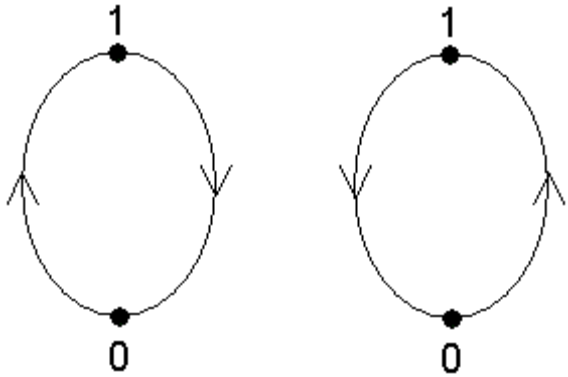

Let us consider a third event r such that $p \leq q \leq p$ and $q \leq r \leq q$. The resulting cycle $p \leq q \leq r \leq q \leq p$ corresponds to four G' graphs, each one beeing the union of two $Z_2$ graphs joining in two nodes. Then the two nodes of the G' graphs are both 2-valued.

$G'_1 = \{0,1\} \cup \{1,0\}$ joining in the nodes (00) and (11)
$G'_2 = \{0,1\} \cup \{0,1\}$ joining in the nodes (01) and (10)
$G'_3 = \{1,0\} \cup \{0,1\}$ joining in the nodes (11) and (00)
$G'_4 = \{1,0\} \cup \{1,0\}$ joining in the nodes (10) and (01).

The above four G' graphs are associated with the four Bell states which form an entangled basis for 2-qubits: $|\Phi_\pm\rangle = \frac{1}{\sqrt{2}}(|11\rangle \pm |00\rangle)$ ; $\quad |\Psi_\pm\rangle = \frac{1}{\sqrt{2}}(|10\rangle \pm |01\rangle)$.

One example of such graphs is given in fig. 3

**Fig. 3**

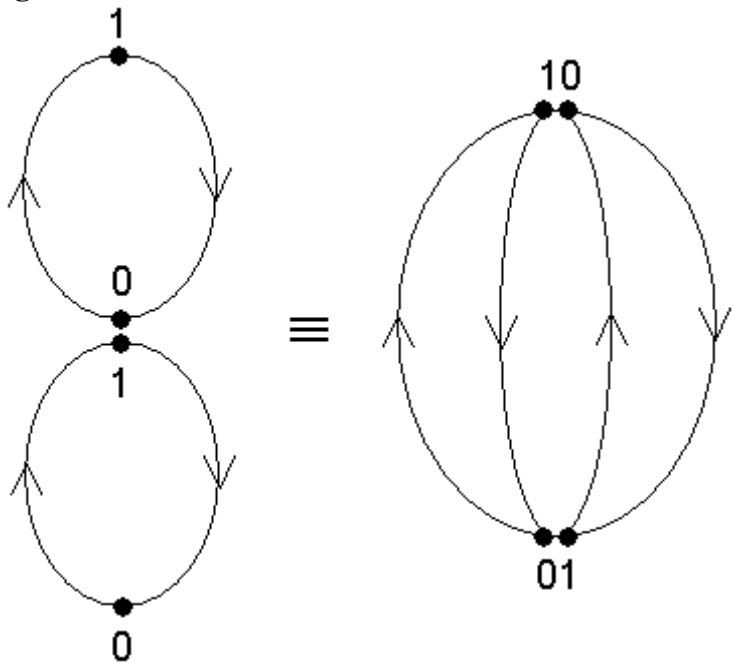

As all the elements of **QC** are related to each other, but not equivalent, all they are entangled qubits, unless the dimension of the discrete space is two, in which case we have only 1-qubit state, as in fig.2.

In general, a cycle in **QC**, with a node of maximal indegree (or outdegree) N, will be associated with N entangled qubits.

The discrete spacetime described by **QC** is then non-causal (because of cyclicity) and non-local (because of entanglement).

**QC** then represents a proto-spacetime endowed with quantum logic, whose events are entangled qubits.

## 5      Unentangled qubits: events of quantum spacetime

Let us now consider the three events p, q and r in the quasi-ordered set **Q**, such that $p \leq q \leq p$ with $p \neq q$ (which is associated with the 1-qubit) and either $q \leq r$ or $r \leq q$ which is associated with the classical bits 0 and 1).

This describes four graphs G' '. Each G' '  has a one 2-valued node and one single- valued node, although both nodes have indegree (or outdegree) 2.

The first G' '  is $Z_2$ graph $\{0,1\}$ with one extra outgoing edge from the node 1, and incoming in the node 0. So, nodes 1 and 0 are identified (1,0). The resulting 2-valued node (1,0) is associated with the state $|10\rangle$.

The second G' '  is $Z_2$ graph $\{0,1\}$ with one extra outgoing edge from the node 0, and incoming in the node 1. The resulting 2-valued node (0,1) will be associated with the state $|01\rangle$.

The third G' '  is $Z_2$ graph $\{0,1\}$ with a loop joining the node 1 to itself. The resulting 2-valued node (1,1) will be associated with the state $|11\rangle$.

The fourth G' ' is $Z_2$ graph with a loop joining the node 0 to itself. The resulting 2-valued node (0,0) is associated with the state $|00\rangle$.

Then the four G' ' graphs provide the unentangled basis $|11\rangle, |00\rangle, |01\rangle, |10\rangle$ for 2-qubits. One example of such G' ' graphs is given in fig. 4.

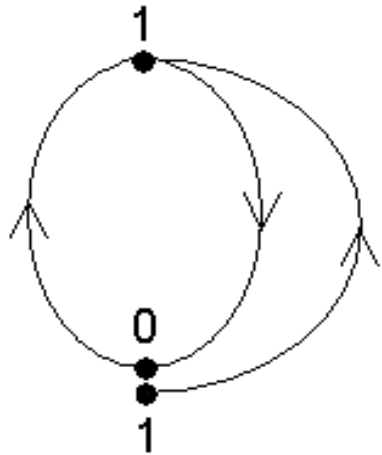

**Fig. 4**

## 6 Information entropy, the arrow of discrete time and micro-causality.

The second law of thermodynamics states that an increase of entropy induces an arrow of time.
An increase of information entropy will also induce an arrow of (discrete) time.
The information entropy of N qubits is $S = N \ln 2$, and if $\Delta S > 0$, we can define an arrow of discrete time $t_N$, where $t_N$ should be directly proportional to some expression of N. However, one should be aware that there are two different situations in QC and in U. In QC, the N qubits are all entangled to each other once for all, and they influence each other simultaneously. The usual causal relation is then meaningless in this case. In fact, there is no increase of information entropy, and an arrow of time cannot be defined. Instead, in U, there are N unentangled qubits, which are product states. For each factor state we can define the information entropies $S_1, S_2 ... S_n ... S_N$ and define an increase of entropy $\Delta S_{n,n'} = (n'-n) \ln 2 > 0$ if n' > n. In this case an arrow of discrete time can be defined.
Let us consider the two elements p and q of Q with $p \leq q \leq p$ and $p \neq q$ (a cycle graph $Z_2$ which is associated with the 1-qubit state). This state is the ground state $|\Psi_0\rangle$ of quantum spacetime, with minimal information entropy $S = \ln 2$ (N=1).
We interpret p and q as virtual events in the time interval $\Delta t = t_P$ where $t_P$ is the Planck time.

The energy associated with this virtual process is: $\Delta E = \frac{\hbar}{\Delta t} = \frac{\hbar}{t_P} = E_P$ where $E_P \approx 10^{19} GeV$ is the Planck energy. This process describes a virtual Planckian black hole whose birth and death are the virtual events p and q. The two virtual events are associated with one $Z_2$ graph. In fact the graph $Z_2$ itself can be considered as the building block of quantum spacetime. In this context, the "event" of quantum spacetime, is the ensemble of two virtual events, and is not a point, but an extended object: a Planckian black hole.

The horizon area of the Planckian black hole is one pixel, i.e., one unit of Planck area $L_P^2$, (where $L_P \approx 10^{-35} m$ is the Planck length), and, in accordance with the quantum version [11] of the holographic principle [14], it encodes one qubit. In conclusion, an event of quantum spacetime, is an extended object which is endowed with the Planck energy, and encodes one unit of quantum information.

Let us now consider the four graphs in section 5, one of which is represented in fig. 4. This is in fact the ensemble of three virtual events which is associated with four $Z_2$ graphs, i.e., with four "events" of quantum spacetime, encoding four (unentangled) qubits.

In general, a number $n_V$ of virtual events (with $n_V = 2,3,4....$) is associated with $N = (n_V - 1)^2$ cycle graphs or "events" of quantum spacetime (with N=1,4,9...) encoding N unentangled qubits $|N\rangle = \frac{1}{\sqrt{2}^N} |Q_1\rangle^{\otimes N}$ where in fact $|Q_1\rangle$ is the ground state $|\Psi_0\rangle$ of quantum spacetime.

As we have seen, the uncertainty in the energy associated with one pair of such virtual events ($n_V = 2$) is the Planck energy. The uncertainty in the energy in a process involving a total number $n_V$ of virtual events, is the Planck energy divided by the total number of pairs $n_V - 1$:

$$\Delta E \equiv E_N = \frac{E_P}{n_V - 1} = \frac{E_P}{\sqrt{N}}.$$

Moreover, the time-energy uncertainty relation should be saturated for every process involving $n_V$ virtual events: $\Delta E \Delta t \equiv E_N t_N \approx \hbar$, from which it follows: $t_N = \sqrt{N} t_P$, which is in fact proportional to (the square root of) the information entropy.

In summary, the information entropy of N events of quantum spacetime induces an arrow of discrete time which is quantized in Planck time units.

The (unentangled) N-qubit states $|N\rangle$ form a causal set $C_N$:

$|N\rangle \leq |M\rangle$ for $t_N \leq t_M$

The $|N\rangle$ states satisfy reflexivity, antisymmetry, and transitivity.

Thus, in $C_N$, micro-causality is recovered.

However, $C_N$ is just a subset of **Q**. In particular, in $C_N$ all entangled states are missing. A similar attempt has been done in terms of evolving spin networks [10]. As the causal set of spin networks and the causal set $C_N$ are strictly related to each other by the holographic principle (see [11] for more details on this), the above arguments hold for the spin networks' case as well.

Let us now look in more detail for the relation between the causal order defined on N unentangled qubits and the causal evolution of spin networks.

Basically, spin networks are graphs embedded in 3-space, with edges labeled by spins j=0, 1/2, 1, 3/2...and vertices labeled by intertwining operators. In loop quantum gravity, spin networks are eigenstates of the area and volume-operators [6]. If a single edge punctures a 2-surface transversely, it contributes an area proportional to: $L_P^2 \sqrt{j(j+1)}$ (where $L_P$ is the Planck length).

The points where the edges end on the surface are called "punctures".

If the surface is punctured in n points, the area is proportional to: $L_P^2 \sum_n \sqrt{j_n(j_n+1)}$.

The cycle graph $Z_2$ in fig. 2 represents a virtual, Planckian black hole, whose surface horizon has an area of one pixel (one unit of Planck area, $L_P^2$), which, by the holographic principle, encodes one qubit. By the above arguments, it follows that there is only one puncture $P_1$ giving rise to one pixel of area, associated with the 1-qubit state. This is the relation between the extended object (the quantum event) and the point $P_1$ (the classical event of a discrete spacetime). In this way, a causal order on N (unentangled) qubits induces causal evolution of spin networks. It should be noticed that the causal evolution of spin networks was originally put "by hand", instead the arrow of discrete time naturally arises in our context as a consequence of the increasing of information entropy.

The situation is schematised in fig. 5, where the surface enclosed by the cycle graph $Z_2$ has an area of one pixel, due to one puncture P of a spin networks' edge.

**Fig. 5**

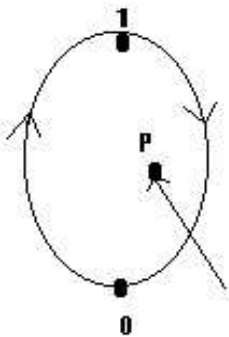